\def\year{2020}\relax
\documentclass[letterpaper]{article} 
\usepackage{aaai20}  
\usepackage{times}  
\usepackage{helvet} 
\usepackage{courier}  
\usepackage[hyphens]{url}  
\usepackage{graphicx} 
\usepackage{graphicx}  
\usepackage{amssymb}
\usepackage{amsmath}
\usepackage{booktabs}
\usepackage{graphics}
\usepackage{multirow}
\usepackage{subfig}

\usepackage[ruled,vlined]{algorithm2e}
\usepackage{makecell}
\usepackage{bbm}
\usepackage{cleveref}
\usepackage{subfloat}
\usepackage{pgfplots}

\crefformat{section}{\S#2#1#3} 
\crefformat{subsection}{\S#2#1#3}
\crefformat{subsubsection}{\S#2#1#3}
\title{Incremental Fairness in Two-Sided Market Platforms: \\On Smoothly Updating Recommendations}
\author{
	Gourab K Patro\\
	Max Planck Inst. for Software Systems, Germany \\
	Indian Institute of Technology Kharagpur, India\\
	\\\And
	Abhijnan Chakraborty\\
	Max Planck Inst. for Software Systems, Germany \\
	\\\AND
	Niloy Ganguly \\
	Indian Institute of Technology Kharagpur, India\\
	\\\And
	Krishna P. Gummadi \\
	Max Planck Inst. for Software Systems, Germany \\
}

\begin{document}
	
	\maketitle
	
	\begin{abstract}
		Major online platforms today can be thought of as two-sided markets with producers and customers of goods and services. 
		There have been concerns that over-emphasis on customer satisfaction by the platforms may affect the well-being of the producers. 
		To counter such issues, few recent works have attempted to incorporate fairness for the producers. 
		However, these studies have overlooked an important issue in such platforms -- to supposedly improve customer utility, 
		the underlying algorithms are frequently updated, causing abrupt changes in the exposure of producers. 
		In this work, we focus on the fairness issues arising out of such frequent updates, and argue for incremental updates of the 
		platform algorithms so that the producers have enough time to adjust (both logistically and mentally) to the change. 
		However, naive incremental updates may become unfair to the customers. Thus focusing on recommendations deployed on two-sided platforms, we formulate an ILP based online optimization to deploy changes incrementally in $\eta$ steps, where we can ensure smooth transition of the exposure of items while 
		guaranteeing a minimum utility for every customer.
		Evaluations over multiple real world datasets show that our proposed mechanism for platform updates can be efficient and fair to both the producers and the customers in two-sided platforms.
	\end{abstract}	
\section{Introduction}
Many popular online platforms today can be thought of as {\it two-sided markets}, such as, sharing economy platforms like Uber, Lyft or Airbnb, e-commerce sites like Amazon, news aggregation services like Google News, location-based review and recommendation services like Yelp, Google Local, employment sites like LinkedIn, Indeed, or hotel aggregators like Booking.com. There are three stakeholders in these markets: (i) {\it producers} of goods and services (e.g., sellers in Amazon, hosts in Airbnb), (ii) {\it customers} who pay for them, and (iii) the {\it platform} 
at the center of the ecosystem. 
Services on these platforms have traditionally been designed to maximize customer satisfaction, since they are the ones directly contributing to the platform revenue, largely ignoring the interest of the other key stakeholder -- \textit{producers}.

Several recent studies have shown how sole focus on the customers may adversely affect the well-being of the producers, as more and more people are depending on two-sided platforms to earn a living~\cite{edelman2017racial,hannak2017bias,abdollahpouri2019multi,chakraborty2017fair,burke2017multisided,graham2017digital}. Subsequently, few research works have attempted to reduce unfairness in these platforms~\cite{suhr2019two,surer2018multistakeholder,geyik2019fairranking}. However, existing works have overlooked an important issue.
They assume that the underlying 
platform algorithms remain unchanged; whereas, to offer supposedly higher customer utility, the platform algorithms go through frequent 
changes and updates. 
Such updates are often very rapid and immediate, leaving no room for the producers to adjust to them. 
For example, with every change in the Facebook Newsfeed curation algorithm, news media outlets (i.e., the producers of news stories) complain about immediate drop in traffic to their websites~\cite{neiman_lab,adexchanger_fb}. 
Similar complaints have been reported also for other two-sided platforms like Amazon~\cite{vox_amazon}.

While maximizing customer's utility may be paramount, we argue that the platform also needs to be fair to the producers while updating. 
Multiple works in behavioral economics have shown that {\it human perceptions of fairness} of a new decision making system 
are influenced by how far the decision outcomes change from the {\it status quo} (i.e., the existing outcomes)~\cite{bediou2014egocentric,kahneman1991anomalies}. 
Motivated by this line of work, in this paper, we propose the notion of {\bf Egocentric Fairness} for the producers, 
which requires that the impact of the changed system should be limited/small. 
We argue that a simple way of being fair would be to implement the change in a phased manner. 
This also has its practical advantage whereby a producer gets time to adjust to the change in demand. 


One naive way of incremental update would be to change the platform for only a subset of customers and then gradually cover everyone. However, such an approach may be unfair to the customers. Since the change is supposed to provide higher satisfaction (utility) for the customers, those who experience the changed platform earlier will get higher utility than the customers covered later. 
To ensure fairness to the customers, we formulate a constrained optimization problem whereby 
at every stage {\bf every customer is guaranteed a minimum utility}, 
while the average change in exposure of the producers should be minimal.
To model this update mechanism, in this paper, we focus on the 
recommendation systems deployed on two-sided platforms (we consider Amazon products and Google Local datasets), 
and consider three common types of updates: 
(i) addition of new features to better estimate customer preferences, (ii) deployment of a new recommendation algorithm reflecting technological advances, or (iii) addition of more data/customer feedback to account for the ever-changing choices of the customers. 

The paper progresses in the following fashion, in~\cref{motivation} we introduce the datasets and update 
types, and perform a detailed experiment to show the impact of immediate update on the producers.
The findings help us to succinctly define fairness from the perspective of producers and customers.
Based upon this understanding of both-sided fairness, the constrained optimization formulation is developed in~\cref{sec:methodology}. 
The formulation takes into consideration several practical details -- for example,
optimization  has to be performed at the level of each customer arrival
and one may or may not have an estimate of the amount of changes which would happen if an update is applied.
The experimental results show that
both efficiency and fairness are ensured to the producers as well as the customers;
the experiments bring forward the lacunae of updating algorithms popular in software engineering domain (used as baselines).  
To our knowledge, this is the first paper which focuses on issues associated with updates on two-sided platforms,
and we believe that this work will be an important addition to the growing literature on fairness of algorithmic decision making systems.
\section{Background and Related work}
\label{related}
{\bf Fairness in Multi-Sided Platforms:}
Recently, few works have looked into the issues of unfairness and biases in platforms with multiple stakeholders.
Disparity in customer utilities has lead to the concerns of both individual and group fairness for customers. 
For example, studies have found instances of {\it group unfairness} -- gender-based discrimination in career ads~\cite{lambrecht2019algorithmic}, or racial bias in guest acceptance by Airbnb hosts~\cite{edelman2017racial}. 
On the other hand, \cite{serbos2017fairness} have looked into {\it individual customer fairness} by 
studying the problem of {\it envy free} tour package recommendations on travel booking sites. 
Similarly, producer fairness relates to the disparity in producer utilities, and touches both group and individual fairness. 
For example,~\cite{hannak2017bias} found racial and gender bias in ratings of freelancers on freelance marketplaces, 
~\cite{chakraborty2019equality} proposed methods to ensure fair representation to different user groups in social media,
~\cite{geyik2019fairranking} proposed fair exposure to candidates from different age and gender groups 
in LinkedIn. 
\cite{biega2018equity} considered individual producer fairness in ranking in gig-economy platforms. 

Few recent works have also explored fairness for both producers and customers.
For example,~\cite{abdollahpouri2019multi,burke2017multisided} categorized different types of multi-stakeholder platforms 
and their desired fairness properties,~\cite{suhr2019two} presented a mechanism for two-sided fairness in matching problems,
~\cite{surer2018multistakeholder} used minimum guarantee constraints for producers 
and diversity constraints for customers while recommending. However, these works have 
assumed that the underlying customer-item relevance model remains unchanged, whereas in reality, the algorithms 
go through frequent updates. 
In this paper, we focus on fairness issues arising out of such 
platform updates in multi-sided platforms. 

\vspace{1mm} 
\noindent{\bf Egocentric Perceptions of Fairness:}
Multiple research works have documented the existence of {\it egocentric biases} in what people perceive as fair. Through experiments in game theory (more specifically, Dictator Games and Ultimatum Games), researchers have observed that individuals take fairness concerns (such as equality) into account while distributing goods among players, and such concerns often originates from one's {\it sense of endowment}~\cite{bediou2014egocentric}. Such endowment effect has also been studied in behavioral economics~\cite{morewedge2015explanations}, where researchers found that individuals perceive a new system to be fair if the new outcomes are similarly beneficial as their {\it status quo} outcomes from the existing system~\cite{kahneman1991anomalies}. Following this line of work, in this paper, we define the notions for {\it egocentric fairness} for producers in two-sided platforms and propose mechanisms to achieve the same.

\vspace{1mm} \noindent
{\bf Incrementalism:}
Incrementalism is a well-studied discipline in public policy making, which advocates for creating policies in iterations where new policy will build upon past policies, incorporating incremental rather than wholesale changes~\cite{hayes1992incrementalism}. Similar to policy issues, 
we argue for incremental algorithmic changes in two-sided platforms to limit large disruptive changes.

\vspace{1mm} \noindent {\bf Minimum Utility Guarantee:}
\cite{pollin2008measure,green2010minimum,falk2006fairness} proposed minimum wage guarantee as a fairness standard, and \cite{lin2016effects,engbom2018earnings} showed evidences of how minimum wage guarantee decreases income inequality.
Inspired by these works, we propose notion of minimum utility guarantee for customer fairness.
\section{Notations and Terminology}
\label{terminology}
In this paper, $U$, $P$, $S$ denote the sets of customers, producers, and items respectively. 
$S_p$ represents the set of all items listed by a producer $p$ such that $\bigcup_{p\in P} S_p=S$.
$R_u$ represents the set of $k$ items recommended to customer $u$; $R_u\subset S$, $|R_u|=k$. We assume $k$ to be the same for every customer. Next, we define the terms used in the paper.

\noindent \textbf{Relevance of Items: }
\label{relevance}
Relevance of an item $s$ to a customer $u$ represents the likelihood that 
$u$ would prefer $s$. 
Formally, we can define relevance as a real function of customer and item;
$V: U \times S \rightarrow \mathbb{R}$, and $V(u,s)$ denotes the relevance of item $s$ to customer $u$.
Alternatively, we can consider $V(u,s)$ as 
the amount of utility gained by customer $u$ if item $s$ is recommended to her.

\noindent \textbf{Customer Utility: }
\label{customer_utility}
The utility of recommendation $R_u$ to $u$ w.r.t. a particular relevance function $V$ can be written as; $\phi(R_u,V)=\sum_{s\in R_u} V(u,s)$.
$u$ will get the maximum possible utility if $k$ most relevant items -- $R_u^*$, is recommended to her; $\phi^\text{max}_u(V)= \phi({R_u^*},V)=\sum_{s\in {R_u^*}}V(u,s)$.
A normalized form of customer utility from a recommendation $R_u$ would be: 
$\phi^\text{norm}(R_u,V)=\frac{\phi({R_u},V)}{\phi^\text{max}_u(V)}=\frac{\phi({R_u},V)}{\phi(R_u^*,V)}$.

\noindent \textbf{Producer Exposure: }
\label{exposure}
The utility of a producer is the total amount of exposure/visibility its items get through recommendations.
The exposure of an item $s$ is the total amount of attention it receives from all the customers to whom $s$ has been recommended.
In an online scenario, $U$ can be thought of as the sequence of customer-visits to the platform  
where 
some customers may visit multiple times. 
If $U([t_1,t_2))$ is the sequence of 
customer-logins in the interval $[t_1,t_2)$, then the exposure of an item $s$ in the same interval will be $E_s([t_1,t_2))=\frac{1}{k}\sum_{u\in U([t_1,t_2))}\mathbbm{1}_{R_u}(s)$, and
that of a producer $p$ will be $E_p([t_1,t_2))=\frac{1}{k}\sum_{s\in S_p}\sum_{u\in U([t_1,t_2))}\mathbbm{1}_{R_u}(s)$\footnote{$\mathbbm{1}_{R_u}(s)$ is $1$ if $s\in R_u$, and $0$ otherwise.}. \\
Note that, in this work, 
we assume that customers pay similar attention to all $k$ recommended items, and leave the consideration 
of position bias (i.e., top-ranked items may get more attention than low-ranked ones) for future work. 
We further assume an one-to-one correspondence between producers and items, and 
{\bf henceforth use the terms `item' and `producer' interchangeably}. This assumption is valid for multiple platforms such as   
restaurant reservation/food delivery (Yelp, Google Local, Uber Eats), freelance marketplaces (Fiverr), human resource matchmaking (LinkedIn, Naukri), and so on. Even for e-commerce platforms where a producer can list multiple items, ensuring fair exposure to individual items would also ensure fairness at the producer level.
Additionally, we have presented a proposal to satisfy fairness specifically at the producer level (not considering at the item level) in the supplementary material.\\
\label{exposure_distribution}
The {\bf distribution of exposure} received by the items can be written as
$D$ = $\{D_s=\frac{E_s}{\sum_{s\in S}E_{s}} \forall s \}$.
Given two exposure distributions $D^\text{old}$ and  $D^\text{new}$, we use L1-norm 
to calculate overall change in exposure: 
$EC$(old,new) = $\sum_{s\in S}|D_s^\text{new}-D_s^\text{old}|$.

\noindent In this paper, we assume that there is no change in overall demand of any item during the update.
Although this assumption may not exactly replicate reality in some situations, but it helps us to focus on the main issue 
of two-sided fairness and bring out the nuances associated with it, rather than the general issue of unpredictability of demand.
\section{Updating Recommendations in \\ Two-Sided Platforms}
\label{motivation}
In this section, we discuss the impact of platform updates on exposure of the producers.
To concretely highlight the impact, we consider certain datasets, as well as different types of updates 
that are undertaken in real-world platforms.

\subsection{Datasets and Types of Updates}
\label{dataset}
In this work, we use the following datasets and test different types of updates on them.

\subsubsection{Amazon Reviews dataset:}
We use the dataset released by 
\cite{he2016ups}, which comprises of customer reviews and ratings 
for different Amazon products 
from the grocery category.
From this dataset, we shortlist $1,000$ most active customers (i.e., who have reviewed most number of products) and $1,000$ most reviewed products, and only consider their corresponding ratings.
Note that the rating act as a proxy to relevance score and the ratings of all the customer-item pairs are not available. 
Data-driven models are used to calculate the missing relevance values of other customer-item pairs.
We test two kinds of updates on this dataset.

\noindent \textbf{\textit{A. Changing the Model}} {\bf (Amazon-M):}
We test updating the recommender system (or the relevance scoring model) from a user-based collaborative filtering~\cite{breese1998empirical} (it works on the assumption that similar users like same set of items) to a more sophisticated latent factorization based model~\cite{koren2009matrix}.

\noindent \textbf{\textit{B. Updating Training Data}} {\bf (Amazon-D):}
The most common type of update is the addition of new training data points. 
Here we calculate the relevance scores using a latent factorization method.
At first the model is trained on the ratings received in the year 2013, and then trained on 2013 and 2014 rating data taken together. 

\noindent We assume that since a platform is adopting a new recommendation algorithm, implicitly that means improved accuracy, otherwise there is no reason for the adoption. As a sanity check, our evaluation on held out ratings data shows improvements of $21.78$\% and $32.46$\% in root-mean-square-error by updating in Amazon-M and Amazon-D repectively.

\subsubsection{Google Local dataset:}
We use data from Google Local, released by 
\cite{he2017translation}, containing data about customers, local businesses and their locations (geographic coordinates), ratings, reviews etc.
At any point in time, we consider each customer's last reviewed location as a proxy for her location. 
We consider all active customers located in New York City and the business entities listed there, 
with more than $10$ reviews. 
The dataset contains $45,305$ customers, $3,029$ businesses and $89,737$ reviews.

\noindent \textbf{\textit{C. Addition of New Feature(s)}} {\bf (GoogleLoc-F):}
Sometimes a new feature (e.g., customers' location) is added to improve the relevance prediction model. We test an update from a purely ratings-based recommendation, $V^\text{old}(u,s)=rating(s)$, to a rating-cum-location based recommendation, $V^\text{new}(u,s)=\frac{rating(s)}{distance(u,s)}$.

\if{0}
As the inter-arrival time distribution of a Poisson process follows an exponential distribution, we sample the inter-arrival times $x$ of customer $u$ using the probability density function $f(x,\lambda_u)=\lambda e^{-\lambda x}$ if $x\geq 0$, and $f(x,\lambda_u)=0$ otherwise.
These inter-arrival times are then mapped onto a unique time axis.
We then use the respective relevance scoring top-$k$ or the proposed approach to find the recommendation set for each of the customer login instances over the time axis.
\fi

\subsection{Modeling Customer Arrivals}
\label{customer_login}
As we do not have the temporal customer arrival/login data, 
we model customer login events as Poisson point processes~\cite{chiu2013stochastic}, where we consider every customer's logins to be independent of each other.
The mean inter-arrival time (time interval between two consecutive arrivals on the platform) of each customer is sampled from a truncated Gaussian distribution (range $[0,2]$) with a mean of $1$ period (exact definition of a period may vary from platform to platform) and variance $0.2$. 

\subsection{Impact of Immediate Updates} 
\label{immediate_update}
\begin{table}[t!]
	\scriptsize
	\begin{center}
		\begin{tabular}{ |c|c c c| } 
			\hline
			\multirow{2}{*}{\bf Dataset} & \multicolumn{3}{c|}{\% of items with change of exposure}\\
			& $< 50\%$ & $50 - 100\%$ & $100+\%$ \\\hline
			{\bf Amazon-M} & 13.7 & 2.5 & 83.8 \\\hline
			{\bf Amazon-D} & 24.1 & 6.7 & 69.2 \\\hline
			{\bf GoogleLoc-F} & 0.12 & 1.17 & 98.71 \\\hline
		\end{tabular}
	\end{center}
	\caption{Percentage change in the exposure of individual items due to immediate update of recommendations.}
	\label{table:impact}
\end{table}

\noindent With the described customer arrival process, we implement updates listed in~\cref{dataset} in an immediate manner and 
report the distribution of the percentage changes in item exposures in Table \ref{table:impact}. 
It is clear from the table that across different types of updates, $69 - 98\%$ of the items experience more than $100$\% change (gain or loss) in their exposure values.
Exposure or visibility often correlates with sales or economic opportunities on which the livelihood of many individuals depends~\cite{wu2009predicting,ghandour2008measuring}.
An abrupt change (loss) in exposure could translate into economic loss or even shutdown; an abrupt gain may lead to degeneration of quality due to demand pressure. 
To capture the unfairness associated with such abrupt changes, we formalize the fairness notions for both producers and customers, as discussed next.

\subsection{Formalizing Fairness in Two-Sided Platforms}
\label{fairness}
\subsubsection{Egocentric Fairness for Producers:}
\label{prod_fairness}
As mentioned earlier, egocentric perception of fairness~\cite{bediou2014egocentric,kahneman1991anomalies} depends on the change from the {\it status quo}.
We define a platform update to be fair to the producers if the difference between the exposure distribution in the new system and that in the old system is minimal.
More formally, if the new and previous exposure distributions are $D^\text{new}$ and $D^\text{old}$ respectively, then a platform is {\it egocentric fair} if  
$\sum_{s \in S} | D_s^\text{new} - D_s^\text{old} | < \epsilon$,
where $\epsilon$ is a small positive number.
\subsubsection{Minimum Guarantee for Customers:}
\label{cust_fairness}
While being fair to the producers, the platform should not compromise on the satisfaction of the customers.
We define a platform to be {\it fair} to the customers if it guarantees a minimum utility for everyone;
{$\phi^\text{norm}(R_u,V^\text{new}) \geq \theta, \forall u\in U$;
where $V^\text{new}$ is the new relevance scoring function to be implemented, and $\theta$ is the utility guarantee provided by the platform\footnote{Our proposal is comparable with the {\it fairness of minimum wage guarantee} (e.g., as required by multiple legislations in US, staring from {\it Fair Labor Standards Act 1938} to {\it Fair Minimum Wage Act 2007})~\cite{pollin2008measure,green2010minimum,falk2006fairness}. While ensuring minimum wage does not itself guarantee equality of income, it has been found to decrease income inequality~\cite{lin2016effects,engbom2018earnings}.}.\\
	
\noindent Table \ref{table:impact} clearly shows that updating recommendations immediately, violates the maxim of {\it egocentric fairness} for the 
producers. 
To ensure fairness, a phased update strategy can be undertaken. This is in line with research works in law, macroeconomics and business philosophy~\cite{malerba1992learning,mintrom1996advocacy,rabin1997fairness}, where they have advocated for {\bf incremental changes} for easy societal adaptation. 
However, updating recommendations incrementally in a two-sided market is challenging due to the dual task of protecting the producers, as well as ensuring a certain level of customer utility. 
We discuss this task in the next section.
\section{Updating Recommendations Incrementally}
\label{sec:methodology}
\begin{figure}[t!]
	\center{
		{\includegraphics[width=0.48\textwidth]{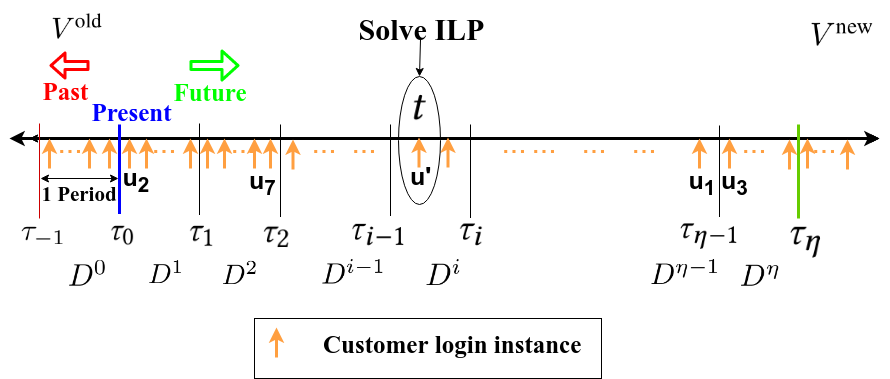}}
	}
	\caption{Timeline representation of incremental update:  
		$V^\text{old}$ and $V^\text{new}$ are the current and new relevance scoring functions.
		The points on time axis are $\tau_{-1}$ ($1$ period in past), $\tau_0$ (present or start time of update), $\tau_{\eta}$ (end time).
		[$\tau_0$,$\tau_\eta$) is the time interval over which an incremental update will be implemented.
		$D^{0}$ is the observed exposure distribution in [$\tau_{-1}$,$\tau_0$).
		$D^{i}$ is the exposure distribution to be observed in step $i$ ([$\tau_{i-1}$,$\tau_{i}$)).
		The objective of the proposed incremental update is to make each of the steps account for small changes in exposure, so that the transition is smooth for the items/producers, 
		while ensuring fair customer utility.}
	\label{fig:timeline}
\end{figure}
\label{methodology}
In this section, we propose to update the recommendations in $\eta$ steps (or $\eta$ periods).
We define points on the time axis like $\tau_{-1}$ ($1$ period in past), $\tau_0$ (present or start point), $\tau_1$, $\cdots$, $\tau_\eta$ (end point) such that [$\tau_{i-1}$,$\tau_i$) represents $i^\text{th}$ period and the targeted update is achieved at $\tau_\eta$. 
In each period, customers visit the platform (as modeled in \cref{customer_login}) and personalized recommendation is provided to each of them.
$D^{0}$ is the observed exposure distribution in [$\tau_{-1}$,$\tau_0$).
Let $D^{i}$ represent the exposure distribution to be observed in step $i$ ([$\tau_{i-1}$,$\tau_{i}$)).
Figure~\ref{fig:timeline} illustrates this online step-wise set up over time.

\subsection{Incremental Update Formulation}
Recommending items to minimize the exposure change can hurt customer utility;
while maximizing customer utility can cause huge changes in exposure.
To address this trade-off, we can formulate optimization problem in two ways:
(i) one where we minimize exposure change constrained to a lower bound on customer utility, and
(ii) another where we maximize customer utility constrained to an upper bound on exposure change.
In this paper, we use the former one since the utility constraints are more interpretable and exposure objectives can be easily operationalized in an online scenario.
We propose to come up with a target exposure distribution {\small$\overline{D^{i}}$} for step $i$, that is,
we try to make the observed exposure distribution in step $i$ as close as possible to {\small$\overline{D^{i}}$}, thus the objective can be written as below.\\
{\small
	\begin{equation}
		\text{minimize} \mathlarger\sum\limits_{s\in S}\bigg|D^i_s-\overline{D^{i}_s}\bigg|\text{,  \space}\forall i\in\{1,\cdots,\eta\}\label{eq:dist_diff}
	\end{equation}}
	This objective needs to be transformed into an online version which deals with each individual customer logging in at certain points of time.
	Assuming a specific customer $u'$ logs into the platform at time $t$ $(t\in [\tau_{i-1},\tau_i))$, the objective transforms into below.
	{\small
		\begin{equation}
			\underset{R_{u'}}{\text{argmin}} \sum\limits_{s\in S}\bigg|E^t_s +\dfrac{\mathbbm{1}_{R_{u'}}(s)}{k} -(|U([\tau_{i-1},t))|+1)\cdot \overline{D^{i}_s}\bigg| \label{eq:item_obj}
		\end{equation}}
		where $E^t_s$ is the exposure of $s$ in $[\tau_{i-1},t)$,
		$\frac{\mathbbm{1}_{R_{u'}}(s)}{k}$ is the attention to $s$ from $u'$,
		$(|U([\tau_{i-1},t))|+1)$ is the total number of customer logins in $[\tau_{i-1},t]$, and
		{\small$\overline{D^{i}_s}$} is the target exposure proportion for $s$ in step $i$.
		As {\small$\overline{D^{i}_s}$} is a fraction, its multiplication with number of customer logins produces the total targeted exposure for $s$ in $[\tau_{i-1},t]$;
		the difference shows how far is the system from the target exposure. 
		
		\noindent{\bf Constraint for Minimum Utility:}
		Along with the above objective, 
		we also have to ensure a minimum utility to the customers, which would be a hard constraint.
		Thus, we use a constraint with lower bound on the normalized customer utility;
		For customer $u'$ at time $t(t\in [\tau_{i-1},\tau_i])$,
		we impose a constraint that the utility (based on new relevance scoring $V^\text{new}$) of the $k$-items chosen for the customer $u'$ in step $i$ must be above a threshold $\theta_i$:
		$\phi^\text{norm}(R_{u'},V^\text{new}) \geq \theta_i$ or
		$\phi(R_{u'},V^\text{new}) \geq (\theta_i\cdot \phi^\text{max}_{u'}(V^\text{new}))$.\\
		\noindent
		We formulate this optimization problem as an Integer Linear Program (ILP).
		For customer $u'$ logging in at time $t$ $(t\in [\tau_{i-1},\tau_i))$, we introduce $|S|$ decision variables: $X_{u',s}$ which is set to $1$ if $s$ is recommended to $u'$, and set to $0$ otherwise. 
		Now we write the ILP as below.
		{\small
			\begin{subequations}
				\begin{equation}
					\underset{X}{\text{argmin}} \mathlarger\sum\limits_{s\in S}\bigg|E^t_s + \dfrac{X_{u',s}}{k} - (|U([\tau_{i-1},t))|+1)\cdot \overline{D^{i}_s}\bigg| \label{eq:ilp_obj}
				\end{equation}
				s.t.
				\begin{equation}
					X_{u',s} \in \{0,1\} \;\; \forall s\in S \label{eq:ilp_const_1}
				\end{equation}
				\begin{equation}
					\sum\limits_{s\in S}X_{u',s}=k \label{eq:ilp_const_2}
				\end{equation}
				\begin{equation}
					\sum\limits_{s\in S}X_{u',s}\cdot V^\text{new}(u',s) \geq (\theta_i\cdot \phi^\text{max}_{u'}(V^\text{new})) \label{eq:ilp_const_3}
				\end{equation}
			\end{subequations}
		}
		Here, constraint-\ref{eq:ilp_const_1} ensures keeping the variables binary.
		Constraint-\ref{eq:ilp_const_2} ensures selecting $k$ items exactly.
		A minimum customer utility is guaranteed by constraint-\ref{eq:ilp_const_3}.
		
		\if{0}
		\noindent The above formulation considers each producer has only a single item listed on the platform, which is true for platforms like Yelp, Google Local, Uber Eats (restaurants are producers), and LinkedIn Jobs (candidates are producers, and recruiters trying to find suitable candidates are the customers).
		In platforms like e-commerce sites, a producer can list many number of items, and fairness would be in smooth transition of producer exposures.
		Ensuring smooth transition at individual items level will also ensure smooth transition for the producers.
		Thus an item-level solution would still work for platforms where each producer can list multiple items.
		However, we also extend our proposition to producer-level in the supplementary material.
		
		The above ILP solves the discussed problem at individual item level.
		This works for platforms like Google local, Yelp, LinkedIn etc., where each listed item (a restaurant, bar, shop or a candidate/person etc.) constitutes one producer.
		However, in platforms like e-commerce sites, a producer can list many number of items, and fairness would relate to producer exposures.
		Ensuring smaller changes for individual items will also ensure small changes in producer exposure.
		However, if a solution is required specifically at the producer-level, then the objective of the ILP can be tweaked as below; \\
		$\underset{X}{\text{argmin}}$ $\mathlarger\sum\limits_{p\in P}\bigg|E^t_p + \mathlarger\sum\limits_{s\in S_p}\frac{ X_{u',s}}{k} - (|U([\tau_{i-1},t))|+1)\cdot \overline{D^{\tau_{i}}_p}\bigg|$\\
		\fi
		
		\subsection{Parameter Setting}
		There are two important parameters in the ILP formualation {\small$\overline{D^{{i}}_s}$} and $\theta_i$ which need to be fixed.

		\subsubsection{Setting {\small$\overline{D^{i}_s}$}:}\label{setting_1}\label{setting_2}
		We propose two different ways to set the target exposure distributions for each step ({\small$\overline{D^i}$} for step $i$).
		
		\noindent \textbf{\textit{A. Estimated Steps:}} Using the current customer arrival frequency (as in $U([\tau_{-1},\tau_0)$) we can find an estimate of the final exposure distribution for the new relevance scoring (i.e., using top-$k$ of $V^\text{new}$ for $U([\tau_{-1},\tau_0)$) and let that be $D^{\text{pred}}$.
		Imagining $D^0$ and $D^\text{pred}$ as points in multidimensional space (with $|S|$ dimensions), our proposition is to enforce certain level of change towards $D^\text{pred}$ in each step.
		Thus we set the target exposure distribution for step $i$ as:
		{\small$\overline{D^i_s}=D^0_s+i\cdot \delta \text{, \space}\forall s,i$},
		where {\small$\delta =\frac{1}{\eta}(D^\text{pred}_s-D^0_s)$}.
		
		\noindent \textbf{\textit{B. Preserving Steps:}}
		Here, we set target exposure distribution of a step to the observed one in last step
		(we try to preserve the observed exposure),
		i.e., {\small$\overline{D^{i}_s}=D^{i-1}_s, \forall s,i$}.
		\subsubsection{Setting $\theta_i$:}\label{theta_setting}
		We use linearly increasing and geometrically increasing settings for $\theta_i$.
		
		\noindent \textbf{\textit{A. Linear Steps:}}
		$\theta_i=\frac{i}{\eta}$, for $1<i<\eta$,
		
		\noindent \textbf{\textit{B. Geometric Steps:}}
		$\theta_i=\theta_{i-1}+2^{-i}$ for $1<i<\eta$, while $\theta_0= 0$, and $\theta_\eta=1$.
		
		\subsection{Approximate Solution with Prefiltering}\label{prefiltering}
		As our ILP operates on the whole item space, huge item space of some systems can be bottleneck for the ILP solvers.
		To deal with this issue, we propose to prefilter the item space, and then run the ILP on filtered (smaller) item space for an approximate solution.
		We prefilter in following two ways and merge the two filtered lists to get a smaller item space:
		(i) top-($k^2$) ($k=$ recommendation set size) items using new relevance scoring ($V^\text{new}$), which can help in satisfying the customer utility constraint, and
		(ii) top-($k^2$) items based on {\small$\big|\frac{E^t_s}{\sum_{s'\in S}E^t_{s'}}- \overline{D^{\tau_i}_s}\big|$}, which can help in minimizing the objective function.
		We test the proposed ILP with both unfiltered and prefiltered item spaces in \cref{experiments}.
		
		\subsection{Baselines}\label{baselines}
		Only a few prior works consider incremental changes; however they do not necessarily cater to two-sided platforms. 
		
		\noindent\textbf{Baseline-1 (CanD): }\label{baseline_canary}\textit{\textbf{Canary deployment}}~\cite{tseitlin2017progressive} (also known as \textit{phased roll out} or \textit{incremental roll out}) is a  popular approach traditionally used in software deployment, where a new software version is slowly rolled out in production for subsets of customers to reduce the risk of imminent failure in an unseen environment. 
		
		\noindent\textbf{Baseline-2 (IRF): }\label{baseline_intermediate}Another approach for incremental update would be to introduce \textbf{\textit{intermediate relevance functions}} for each of the steps 
		(gradually moving the relevance scores from $V^\text{old}$ to $V^\text{new}$); relevance function for step $i$: {\small$V^i(u,s)=(1-\frac{i}{\eta})\cdot V^\text{old}(u,s)+\frac{i}{\eta}\cdot V^\text{new}(u,s), \forall u, s$}.
		We can recommend the top-$k$ according to $V^i(u,s)$ in step $i$.
\section{Experimental Evaluation}\label{experiments}
For each customer $u'$ logging into the platform at time $t$, we solve the proposed ILP with different settings
which gives a set of items to be recommended.
Using these results, we calculate and record item exposures and customer utilities in each step.
We set the number of steps $\eta=10$, and size of recommendation $k=10$.
We use \textit{cvxpy} ({\tt cvxpy.org}) paired with \textit{Gurobi solver} ({\tt gurobi.com}) for solving the ILP. 
In this section, we use the following abbreviations: {\bf E- Estimated, P- Preserving steps in} {\small$\overline{D^i}$}; {\bf L- Linear, G- Geometric steps in} $\theta_i$; {\bf PF- Prefiltering}.

\begin{table*}[t!]
	\tiny
	\begin{center}
		\begin{tabular}{|l|c c c|c c c|c c c|}
			\hline
			\multirow{2}{*}{\bf Method} &  \multicolumn{3}{c|}{\makecell{{\bf Amazon-M}\\($EC(0,\eta)=1.922$)}} & \multicolumn{3}{c|}{\makecell{{\bf Amazon-D}\\($EC(0,\eta)=1.557$)}} & \multicolumn{3}{c|}{\makecell{{\bf GoogleLoc-F}\\($EC(0,\eta)=1.967$)}} \\\cline{2-10}
			{} & $\Upsilon$ & $\pi$ & $Z$ & $\Upsilon$ & $\pi$ & $Z$ & $\Upsilon$ & $\pi$ & $Z$  \\\hline
			{\bf CanD} & $1.63$ & $0.17$ & $0.99$ & $1.34$ & $0.21$ & $0.99$ & $1.38$ & $0.16$ & $0.99$\\\hline
			{\bf IRF} & $1.48$ & $0.95$ & $0.62$ & $1.46$ & $0.19$ & $0.98$ & $1.46$ & $0.98$ & $0.59$\\\hline
			{\bf ILP-EL} & $\pmb{1.03}$ & $\pmb{0.12}$ & $\pmb{0.99}$ & $\pmb{1.04}$ & $\pmb{0.13}$ & $\pmb{0.99}$ & $\pmb{1.04}$ & $\pmb{0.11}$ & $\pmb{0.99}$\\\hline
			{\bf ILP-EL}(PF) & $\pmb{1.16}$ & $\pmb{0.13}$ & $\pmb{0.99}$ & $\pmb{1.45}$ & $\pmb{0.16}$ & $\pmb{0.99}$ & $\pmb{1.33}$ & $\pmb{0.15}$ & $\pmb{0.99}$\\\hline
			{\bf ILP-EG} & $1.18$ & $0.28$ & $0.93$ & $1.12$ & $0.22$ & $0.97$ & $1.16$ & $0.19$ & $0.96$\\\hline
			{\bf ILP-PL} & $1.06$ & $0.46$ & $0.61$ & $1.02$ & $0.74$ & $0.33$ & $1.08$ & $0.29$ & $0.90$\\\hline
			{\bf ILP-PG} & $1.10$ & $0.30$ & $0.86$ & $1.08$ & $0.23$ & $0.86$ & $1.24$ & $0.29$ & $0.90$\\\hline
		\end{tabular}
	\end{center}
	\caption{Producer-Centric Metrics for $\eta=10$ and $k=10$. CanD and IRF are the two baselines. For ILP-based methods we use abbreviations like; E: Estimated, P: Preserving steps in {\small$\overline{D^i}$}; L: Linear, G: Geometric steps in $\theta_i$ ; PF: Prefiltering item space.}
	\label{table:producer_metrics}
\end{table*}

\subsection{Producer-Centric Metrics}
First we define metric for change in exposure.

\noindent{\bf Exposure Change ($EC$):} Given two exposure distributions $D^\text{old}$ and  $D^\text{new}$, exposure change ($EC$) is given by their L1 distance (also defined in \cref{exposure_distribution});
{\small $EC(\text{old},\text{new}) =  \sum_{s\in S}|D_s^\text{new}-D_s^\text{old}|$}
	Based on $EC$, we define the following three metrics. 
	
	\noindent \textbf{\textit{A. Transition Path Length ($\Upsilon$)}} - Efficiency metric. \\
	It is the sum of all exposure changes that the transition has gone through, relative to that of an immediate update.
	{\small{
			\begin{equation}
				\Upsilon=  \frac{\sum_{i\in \{1\cdots \eta\}}EC(i-1,i)}{EC(0,\eta)}
			\end{equation}
			The {\bf lower} the path length, the {\bf more efficient} is the transition.
		}}
		
		\noindent \textbf{\textit{B. Maximum Transition Cost ($\Pi$)}} - Fairness metric.\\ 
		{\small{
				\begin{equation}
					\Pi=\dfrac{\underset{i}{\text{max}} [EC(i-1,i)]}{EC(0,\eta)}
				\end{equation}
			}}
			$\Pi$ checks the largest change during the incremental transition process relative to that of an immediate update; 
			even a single big exposure change is undesirable as it is inherently disadvantageous (unfair) for producers.
			
			\noindent \textbf{\textit{C. Transition Inequality ($Z$)}} - Fairness metric. \\
			Transition Inequality captures the dissimilarity in the quantum of transition among the steps, measured by entropy as defined below.
			{\small{
					\begin{equation}
						Z=-\sum_{i\in \{1\cdots \eta\}}(\frac{EC(i-1,i)}{M})\log_{10}(\frac{EC(i-1,i)}{M})
					\end{equation}
					where $M =\sum_{i\in \{1\cdots \eta\}}EC(i-1,i)$.
				}}\\
				An {\bf ideal update} will have high efficiency (or {\bf low} $\pmb{\Upsilon}$), and small \& equal sized changes (or {\bf low} $\pmb{\pi}$ and {\bf high} $\pmb{Z}$). 
				
\newenvironment{customlegend}[1][]{%
	\begingroup
	\csname pgfplots@init@cleared@structures\endcsname
	\pgfplotsset{#1}%
}{%
\csname pgfplots@createlegend\endcsname
\endgroup
}%
\def\addlegendimage{\csname pgfplots@addlegendimage\endcsname}
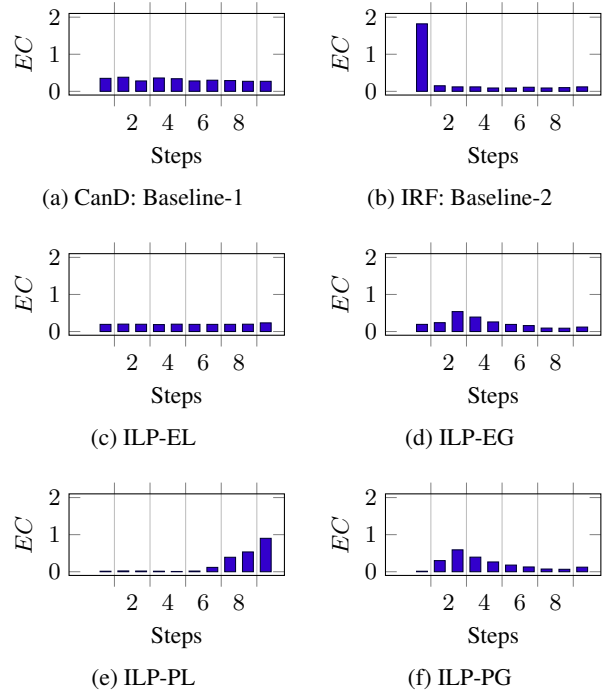
\begin{figure}[t!]
	\small
	\subfloat[{CanD: Baseline-1}]{\pgfplotsset{width=0.25\textwidth,compat=1.9}
		\begin{tikzpicture}
		\begin{axis}[
		width=0.25\textwidth,
		height=0.15\textwidth,
		xlabel=Steps,
		ylabel= $EC$,
		enlargelimits=0.05,
		ybar interval=0.61,
		xmin=0, xmax=11,
		ymin=0, ymax=2,
		xtick={2,4,6,8,10},
		ytick={0,1,2},
		]
		\addplot[blue!20!black,fill=blue!80!red] 
		coordinates {(1,0.35)(2,0.38)(3,0.28)(4,0.36)(5,0.34)(6,0.28)(7,0.30)(8,0.29)(9,0.27)(10,0.27)(11,0)};
		\end{axis}
		\end{tikzpicture}}
	\hfil
	\subfloat[{IRF: Baseline-2}]{\pgfplotsset{width=0.25\textwidth,compat=1.9}
		\begin{tikzpicture}
		\begin{axis}[
		width=0.25\textwidth,
		height=0.15\textwidth,
		xlabel=Steps,
		ylabel= $EC$,
		enlargelimits=0.05,
		ybar interval=0.61,
		xmin=0, xmax=11,
		ymin=0, ymax=2,
		xtick={2,4,6,8,10},
		ytick={0,1,2},
		]
		\addplot[blue!20!black,fill=blue!80!red] 
		coordinates {(1,1.82)(2,0.15)(3,0.12)(4,0.12)(5,0.09)(6,0.09)(7,0.11)(8,0.09)(9,0.10)(10,0.12)(11,0)};
		
		\end{axis}
		\end{tikzpicture}}
	\vfil
	
	\subfloat[{ILP-EL}]{\pgfplotsset{width=0.25\textwidth,compat=1.9}
		\begin{tikzpicture}
		\begin{axis}[
		width=0.25\textwidth,
		height=0.15\textwidth,
		xlabel=Steps,
		ylabel= $EC$,
		enlargelimits=0.05,
		ybar interval=0.61,
		xmin=0, xmax=11,
		ymin=0, ymax=2,
		xtick={2,4,6,8,10},
		ytick={0,1,2},
		]
		
		\addplot[blue!20!black,fill=blue!80!red] 
		coordinates {(1, 0.1928205854045175)
			(2, 0.19953622656138398)
			(3, 0.19595257111517536)
			(4, 0.1873239602978064)
			(5, 0.19789564301524187)
			(6, 0.19311078646657553)
			(7, 0.19281778784381204)
			(8, 0.19419648411809648)
			(9, 0.19740289069557318)
			(10, 0.23340936384921698)(11,0)};
		
		\end{axis}
		\end{tikzpicture}}
	\hfil
	\subfloat[{ILP-EG}]{\pgfplotsset{width=0.25\textwidth,compat=1.9}
		\begin{tikzpicture}
		\begin{axis}[
		width=0.25\textwidth,
		height=0.15\textwidth,
		xlabel=Steps,
		ylabel= $EC$,
		enlargelimits=0.05,
		ybar interval=0.61,
		xmin=0, xmax=11,
		ymin=0, ymax=2,
		xtick={2,4,6,8,10},
		ytick={0,1,2},
		]
		
		\addplot[blue!20!black,fill=blue!80!red] 
		coordinates {(1, 0.19263884957255237)
			(2, 0.23885658321601255)
			(3, 0.5375185334248123)
			(4, 0.389933501580765)
			(5, 0.258519157635362)
			(6, 0.1912696267364008)
			(7, 0.1604823815020026)
			(8, 0.09291728138312627)
			(9, 0.08959848809014347)
			(10, 0.12001244135714428)(11,0)};
		
		\end{axis}
		\end{tikzpicture}}
	\vfil
	\subfloat[{ILP-PL}]{\pgfplotsset{width=0.25\textwidth,compat=1.9}
		\begin{tikzpicture}
		\begin{axis}[
		width=0.25\textwidth,
		height=0.15\textwidth,
		xlabel=Steps,
		ylabel= $EC$,
		enlargelimits=0.05,
		ybar interval=0.61,
		xmin=0, xmax=11,
		ymin=0, ymax=2,
		xtick={2,4,6,8,10},
		ytick={0,1,2},
		]
		\addplot[blue!20!black,fill=blue!80!red] 
		coordinates {(1, 0.01320689199645207)
			(2, 0.021185244128868977)
			(3, 0.018737579216706053)
			(4, 0.01374300664196762)
			(5, 0.006756523040270203)
			(6, 0.0180094701285201)
			(7, 0.11682351635747823)
			(8, 0.3896816495427187)
			(9, 0.5335140006052184)
			(10, 0.9007311457280328)(11,0)};
		
		\end{axis}
		\end{tikzpicture}}
	\hfil
	\subfloat[{ILP-PG}]{\pgfplotsset{width=0.25\textwidth,compat=1.9}
		\begin{tikzpicture}
		\begin{axis}[
		width=0.25\textwidth,
		height=0.15\textwidth,
		xlabel=Steps,
		ylabel= $EC$,
		enlargelimits=0.05,
		ybar interval=0.61,
		xmin=0, xmax=11,
		ymin=0, ymax=2,
		xtick={2,4,6,8,10},
		ytick={0,1,2},
		]
		\addplot[blue!20!black,fill=blue!80!red] 
		coordinates {(1, 0.013206891996452072)
			(2, 0.2999793923813025)
			(3, 0.5906118167656871)
			(4, 0.39306668346516294)
			(5, 0.26349176711909694)
			(6, 0.1790604273131962)
			(7, 0.12849811125087904)
			(8, 0.07263904441955944)
			(9, 0.06556363809252125)
			(10, 0.12257220739051139)(11,0)};	
		
		\end{axis}
		\end{tikzpicture}}
	\hfil
	\caption{Exposure changes at all the steps ($EC(i-1,i)=\sum_{s\in S}|D^i_s-D^{i-1}_s|$) are plotted (x-axis: steps $i$, y-axis: $EC$) for {\bf Amazon-M}. Hyperparameters: $\eta=10$, $k=10$.}\label{fig:exp_change_amazon1}
	
\end{figure}

\begin{figure}[t!]
	\small
	\center{
		\subfloat{\pgfplotsset{width=0.22\textwidth,compat=1.9}
			\begin{tikzpicture}
			\begin{axis}[		
			title={Amazon-M},
			xlabel={Steps},
			ylabel={mean $\phi$},
			xmin=0, xmax=10,
			ymin=0.5, ymax=1,
			xtick={0,1,2,3,4,5,6,7,8,9,10},
			ytick={0,0.2,0.4,0.6,0.8,1.0},
			ymajorgrids=true,
			grid style=dashed,
			]
			\addplot[
			color=red,
			mark=triangle,
			]
			coordinates {
				(0,0.57)(1,0.6)(2,0.7)(3,0.75)(4,0.78)(5,0.82)(6,0.87)(7,0.91)(8,0.95)(9,0.99)(10,1)};
			
			\addplot[
			color=blue,
			mark=o,
			]
			coordinates {(0,0.57)(1,0.97)(2,0.99)(3,0.99)(4,0.99)(5,0.99)(6,0.99)(7,0.99)(8,0.99)(9,1)(10,1)};
			
			\addplot[
			color=black,
			mark=square,
			]
			coordinates {(0,0.57)(1,0.6)(2,0.63)(3,0.66)(4,0.69)(5,0.72)(6,0.75)(7,0.78)(8,0.81)(9,0.9)(10,1)};
			
			\addplot[
			color=green,
			mark=diamond,
			]
			coordinates {(0,0.57)(1,0.6)(2,0.75)(3,0.87)(4,0.94)(5,0.97)(6,1)(7,1)(8,1)(9,1)(10,1)};
			
			\addplot[
			color=orange,
			mark=x,
			]
			coordinates {(0,0.57)(1,0.57)(2,0.57)(3,0.57)(4,0.57)(5,0.57)(6,0.64)(7,0.73)(8,0.81)(9,0.9)(10,1)};
			
			\addplot[
			color=magenta,
			mark=,
			]
			coordinates {(0,0.57)(1,0.57)(2,0.78)(3,0.87)(4,0.94)(5,0.97)(6,1)(7,1)(8,1)(9,1)(10,1)};
			
			\end{axis}
			\end{tikzpicture}}
		\hfil
		\subfloat{\pgfplotsset{width=0.22\textwidth,compat=1.9}
			\begin{tikzpicture}
			\begin{axis}[
			title={Amazon-M},
			xlabel={Steps},
			ylabel={std $\phi$},
			xmin=0, xmax=10,
			ymin=0, ymax=0.25,
			xtick={0,1,2,3,4,5,6,7,8,9,10},
			ytick={0,0.05,0.10,0.15,0.20,0.25},
			scaled y ticks=false,
			yticklabel style={/pgf/number format/fixed},
			ymajorgrids=true,
			grid style=dashed,
			]
			\addplot[
			color=red,
			mark=triangle,
			]
			coordinates {(0, 0.04406078729398601)
				(1, 0.14445099041471895)
				(2, 0.17556068236065878)
				(3, 0.20339870794685447)
				(4, 0.20978852805855344)
				(5, 0.21701296786544824)
				(6, 0.21096865579812893)
				(7, 0.2009833274628267)
				(8, 0.16794696428422506)
				(9, 0.13288681289167037)
				(10, 0.024870011363602357)
			};
			
			\addplot[
			color=blue,
			mark=o,
			]
			coordinates {(0, 0.04406078729398601)
				(1, 0.02975238012572551)
				(2, 0.01312803433854977)
				(3, 0.00480265830363329)
				(4, 0.001340229625458171)
				(5, 0.0003583171968968401)
				(6, 2.327551854917755e-05)
				(7, 4.4510293242939486e-07)
				(8, 3.207156224383236e-05)
				(9, 7.91402982747437e-18)
				(10, 0.0)
			};
			
			\addplot[
			color=black,
			mark=square,
			]
			coordinates {(0, 0.04406078729398601)
				(1, 0.055605991922864556)
				(2, 0.06436853514839742)
				(3, 0.06555574532221337)
				(4, 0.0665742279462516)
				(5, 0.06915197581139057)
				(6, 0.06716953632153093)
				(7, 0.05922932359491517)
				(8, 0.03965532924244906)
				(9, 0.004101281516931763)
				(10, 1.7334532154367282e-16)};
			
			\addplot[
			color=green,
			mark=diamond,
			]
			coordinates {(0, 0.04406078729398601)
				(1, 0.05467883389174801)
				(2, 0.002996707714395536)
				(3, 0.0013698133963485695)
				(4, 0.001679577151661124)
				(5, 0.0017998571070349878)
				(6, 0.002497766831963303)
				(7, 0.003309776982995917)
				(8, 0.0015778720907293849)
				(9, 0.0007308702784981613)
				(10, 1.7334532154367282e-16)
			};
			
			\addplot[
			color=orange,
			mark=x,
			]
			coordinates {(0, 0.04406078729398601)
				(1, 0.04547322880240116)
				(2, 0.04414235671454582)
				(3, 0.046135850330007826)
				(4, 0.04546911195197136)
				(5, 0.04269763196092303)
				(6, 0.007224313230665687)
				(7, 0.001765258015317472)
				(8, 0.0013716396797292937)
				(9, 0.0015747152600923582)
				(10, 1.7334532154367282e-16)
			};
			
			\addplot[
			color=magenta,
			mark=,
			]
			coordinates {(0, 0.04406078729398601)
				(1, 0.04406078729398601)
				(2, 0.001435871704012756)
				(3, 0.0015503445363209381)
				(4, 0.0016441390171788685)
				(5, 0.0017020883159361432)
				(6, 0.001437294302992591)
				(7, 0.001782523191035288)
				(8, 0.0012512649011723286)
				(9, 0.0007325248374904747)
				(10, 1.7334532154367282e-16)};
			
			\end{axis}
			\end{tikzpicture}}
		\vfil
		\subfloat{\pgfplotsset{width=0.22\textwidth,compat=1.9}
			\begin{tikzpicture}
			\begin{axis}[
			title={Amazon-D},
			xlabel={Steps},
			ylabel={mean $\phi$},
			xmin=0, xmax=10,
			ymin=0.5, ymax=1,
			xtick={0,1,2,3,4,5,6,7,8,9,10},
			ytick={0,0.2,0.4,0.6,0.8,1.0},
			ymajorgrids=true,
			grid style=dashed,
			]
			\addplot[
			color=red,
			mark=triangle,
			]
			coordinates {
				(0,0.61)(1,0.77)(2,0.8)(3,0.82)(4,0.84)(5,0.87)(6,0.89)(7,0.92)(8,0.95)(9,0.97)(10,1)};
			
			\addplot[
			color=blue,
			mark=o,
			]
			coordinates {(0,0.61)(1,0.76)(2,0.82)(3,0.87)(4,0.91)(5,0.95)(6,0.97)(7,0.98)(8,0.99)(9,0.99)(10,1)};
			
			\addplot[
			color=black,
			mark=square,
			]
			coordinates {(0,0.61)(1,0.61)(2,0.64)(3,0.67)(4,0.7)(5,0.73)(6,0.76)(7,0.79)(8,0.82)(9,0.9)(10,1)};
			
			\addplot[
			color=green,
			mark=diamond,
			]
			coordinates {(0,0.61)(1,0.61)(2,0.75)(3,0.89)(4,0.95)(5,0.97)(6,1)(7,1)(8,1)(9,1)(10,1)};
			
			\addplot[
			color=orange,
			mark=x,
			]
			coordinates {(0,0.61)(1,0.73)(2,0.73)(3,0.73)(4,0.73)(5,0.73)(6,0.73)(7,0.73)(8,0.8)(9,0.9)(10,1)};
			
			\addplot[
			color=magenta,
			mark=,
			]
			coordinates {(0,0.61)(1,0.73)(2,0.77)(3,0.88)(4,0.93)(5,0.96)(6,0.98)(7,1)(8,1)(9,1)(10,1)};
			
			\end{axis}
			\end{tikzpicture}}
		\hfil
		\subfloat{\pgfplotsset{width=0.22\textwidth,compat=1.9}
			\begin{tikzpicture}
			\begin{axis}[
			title={Amazon-D},
			xlabel={Steps},
			ylabel={std $\phi$},
			xmin=0, xmax=10,
			ymin=0, ymax=0.15,
			xtick={0,1,2,3,4,5,6,7,8,9,10},
			ytick={0,0.05,0.1,0.15},
			scaled y ticks=false,
			yticklabel style={/pgf/number format/fixed},
			ymajorgrids=true,
			grid style=dashed,
			]
			\addplot[
			color=red,
			mark=triangle,
			]
			coordinates {(0, 0.04278140477516436)
				(1, 0.10706355794114507)
				(2, 0.12152468620979849)
				(3, 0.13429996799680777)
				(4, 0.13585208120784154)
				(5, 0.13802505196691428)
				(6, 0.13347786915723772)
				(7, 0.12497898919961381)
				(8, 0.10874636392650504)
				(9, 0.07696230335938932)
				(10, 0)
			};
			
			\addplot[
			color=blue,
			mark=o,
			]
			coordinates {(0, 0.04278140477516436)
				(1, 0.043331013282681144)
				(2, 0.040847024404838714)
				(3, 0.03863435300428974)
				(4, 0.02879918393486895)
				(5, 0.020967476719875297)
				(6, 0.01316663867493766)
				(7, 0.00866707569358935)
				(8, 0.0042759599659749235)
				(9, 0.0013664914743530543)
				(10, 0.0)
			};
			
			\addplot[
			color=black,
			mark=square,
			]
			coordinates {(0, 0.04278140477516436)
				(1, 0.046531072637978274)
				(2, 0.044432736408544504)
				(3, 0.04827355037046346)
				(4, 0.04845286218878835)
				(5, 0.048288932158743136)
				(6, 0.0468306588559575)
				(7, 0.04662125549960067)
				(8, 0.04057328897872325)
				(9, 0.007982999813947664)
				(10, 1.559271568489262e-16)
			};
			
			\addplot[
			color=green,
			mark=diamond,
			]
			coordinates {(0, 0.04278140477516436)
				(1, 0.046694477352145214)
				(2, 0.025126436434122705)
				(3, 0.0014035611643548396)
				(4, 0.0013856051811154536)
				(5, 0.0015097138787352558)
				(6, 0.0014000968290054522)
				(7, 0.0018697953455435001)
				(8, 0.0014976697319770268)
				(9, 0.0007451134625632112)
				(10, 1.559271568489262e-16)
			};
			
			\addplot[
			color=orange,
			mark=x,
			]
			coordinates {(0, 0.04278140477516436)
				(1, 0.04278140477516436)
				(2, 0.0436308424765259)
				(3, 0.04327070053239998)
				(4, 0.0427018259703138)
				(5, 0.04268924706376783)
				(6, 0.042878178359684584)
				(7, 0.03181465622202689)
				(8, 0.0017525926365839178)
				(9, 0.0015397170619975292)
				(10, 1.559271568489262e-16)
			};
			
			\addplot[
			color=magenta,
			mark=,
			]
			coordinates {(0, 0.04278140477516436)
				(1, 0.04292342153327445)
				(2, 0.007975855838969181)
				(3, 0.001426538468158916)
				(4, 0.0018343543302111504)
				(5, 0.0019245515436819025)
				(6, 0.001495888739904861)
				(7, 0.001479765773354502)
				(8, 0.001344225226696581)
				(9, 0.0007533224021059013)
				(10, 1.559271568489262e-16)
			};
			
			\end{axis}
			\end{tikzpicture}}
		\vfil
		\subfloat{\pgfplotsset{width=0.22\textwidth,compat=1.9}
			\begin{tikzpicture}
			\begin{axis}[
			title={GoogleLoc-F},
			xlabel={Steps},
			ylabel={mean $\phi$},
			xmin=0, xmax=10,
			ymin=0, ymax=1,
			xtick={0,1,2,3,4,5,6,7,8,9,10},
			ytick={0,0.2,0.4,0.6,0.8,1.0},
			ymajorgrids=true,
			grid style=dashed,
			]
			\addplot[
			color=red,
			mark=triangle,
			]
			coordinates {
				(0,0.19)(1,0.22)(2,0.35)(3,0.42)(4,0.53)(5,0.61)(6,0.73)(7,0.82)(8,0.9)(9,0.95)(10,1)};
			
			\addplot[
			color=blue,
			mark=o,
			]
			coordinates {(0,0.19)(1,0.98)(2,0.99)(3,0.99)(4,0.99)(5,0.99)(6,0.99)(7,0.99)(8,0.99)(9,0.99)(10,1)};
			
			\addplot[
			color=black,
			mark=square,
			]
			coordinates {(0,0.19)(1,0.19)(2,0.27)(3,0.36)(4,0.45)(5,0.54)(6,0.63)(7,0.72)(8,0.81)(9,0.92)(10,1)};
			
			\addplot[
			color=green,
			mark=diamond,
			]
			coordinates {(0,0.19)(1,0.52)(2,0.76)(3,0.87)(4,0.93)(5,0.95)(6,0.98)(7,1)(8,1)(9,1)(10,1)};
			
			\addplot[
			color=orange,
			mark=x,
			]
			coordinates {(0,0.19)(1,0.2)(2,0.33)(3,0.4)(4,0.5)(5,0.58)(6,0.67)(7,0.75)(8,0.83)(9,0.92)(10,1)};
			
			\addplot[
			color=magenta,
			mark=,
			]
			coordinates {(0,0.19)(1,0.53)(2,0.77)(3,0.87)(4,0.94)(5,0.97)(6,1)(7,1)(8,1)(9,1)(10,1)};
			
			\end{axis}
			\end{tikzpicture}}
		\hfil
		\subfloat{\pgfplotsset{width=0.22\textwidth,compat=1.9}
			\begin{tikzpicture}
			\begin{axis}[
			title={GoogleLoc-F},
			xlabel={Steps},
			ylabel={std $\phi$},
			xmin=0, xmax=10,
			ymin=0, ymax=0.25,
			xtick={0,1,2,3,4,5,6,7,8,9,10},
			ytick={0,0.1,0.1,0.25},
			ymajorgrids=true,
			grid style=dashed,
			]
			\addplot[
			color=red,
			mark=triangle,
			]
			coordinates {(0,0.12239897473318741)
				(1, 0.2280220632724822)
				(2, 0.21989106769925576)
				(3, 0.20392604868226677)
				(4, 0.19032555501465134)
				(5, 0.1709266815314203)
				(6, 0.15941156958120545)
				(7, 0.1399468684374364)
				(8, 0.10970874591612036)
				(9, 0.08105512003107462)
				(10, 0.0159248913201684)
			};
			
			\addplot[
			color=blue,
			mark=o,
			]
			coordinates {(0,0.12239897473318741)
				(1, 0.022285734406774668)
				(2, 0.006792422730809915)
				(3, 0.0041074260018691185)
				(4, 0.0018050766387186442)
				(5, 0.0007759335510034296)
				(6, 0.0003630882106303501)
				(7, 0.0002111219245903238)
				(8, 0.00012595697761109094)
				(9, 1.789851019438273e-05)
				(10, 0.0)
			};
			
			\addplot[
			color=black,
			mark=square,
			]
			coordinates {(0,0.12239897473318741)
				(1, 0.12239897473318741)
				(2, 0.1216208127464821)
				(3, 0.09869852492408068)
				(4, 0.06751182730367397)
				(5, 0.04561613407204164)
				(6, 0.027795063286287225)
				(7, 0.01423393943995984)
				(8, 0.008388174998557143)
				(9, 0.006596903603617976)
				(10, 1.698332204945911e-16)};
			
			\addplot[
			color=green,
			mark=diamond,
			]
			coordinates {(0,0.12239897473318741)
				(1, 0.05073883672912141)
				(2, 0.02181781816766016)
				(3, 0.015553673946833532)
				(4, 0.013678172504503946)
				(5, 0.013019740514512474)
				(6, 0.00523691714200083)
				(7, 0.001864058430514514)
				(8, 0.0007765679924660305)
				(9, 0.0003560532994423128)
				(10, 1.698332204945911e-16)};
			
			\addplot[
			color=orange,
			mark=x,
			]
			coordinates {(0,0.12239897473318741)
				(1, 0.12239897473318741)
				(2, 0.1216208127464821)
				(3, 0.09869852492408068)
				(4, 0.06751182730367397)
				(5, 0.04561613407204164)
				(6, 0.027795063286287225)
				(7, 0.01423393943995984)
				(8, 0.008388174998557143)
				(9, 0.006596903603617976)
				(10, 1.698332204945911e-16)};
			
			\addplot[
			color=magenta,
			mark=,
			]
			coordinates {(0,0.12239897473318741)
				(1, 0.05117656669900664)
				(2, 0.021269433701695324)
				(3, 0.013724298673582042)
				(4, 0.010713366910868515)
				(5, 0.010447533147136413)
				(6, 0.006031991988006218)
				(7, 0.0029295228200121193)
				(8, 0.0011872647218363862)
				(9, 0.00044792367966145166)
				(10, 2.996478895397076e-16)
			};
			
			\end{axis}
			\end{tikzpicture}}
		\vfil
		\subfloat{\pgfplotsset{width=.5\textwidth,compat=1.9}
			\begin{tikzpicture}
			\begin{customlegend}[legend entries={{CanD  },{IRF },{ILP-EL  },{ILP-EG  },{ILP-PL  },{ILP-PG  }},legend columns=3]
			\addlegendimage{red,mark=triangle,sharp plot}
			\addlegendimage{blue,mark=o,sharp plot}
			\addlegendimage{black,mark=square,sharp plot}
			\addlegendimage{green,mark=diamond,sharp plot}
			\addlegendimage{orange,mark=x,sharp plot}
			\addlegendimage{magenta,mark=,sharp plot}
			\end{customlegend}
			\end{tikzpicture}}
	}
	\caption{Mean (left column) and standard deviation (right) of $\phi^\text{norm}(R_u,V^\text{new})$ 
		$\forall u\in U([\tau_{i-1},\tau_i))$ for each step $i$.}
	\label{fig:utility_main}
\end{figure}
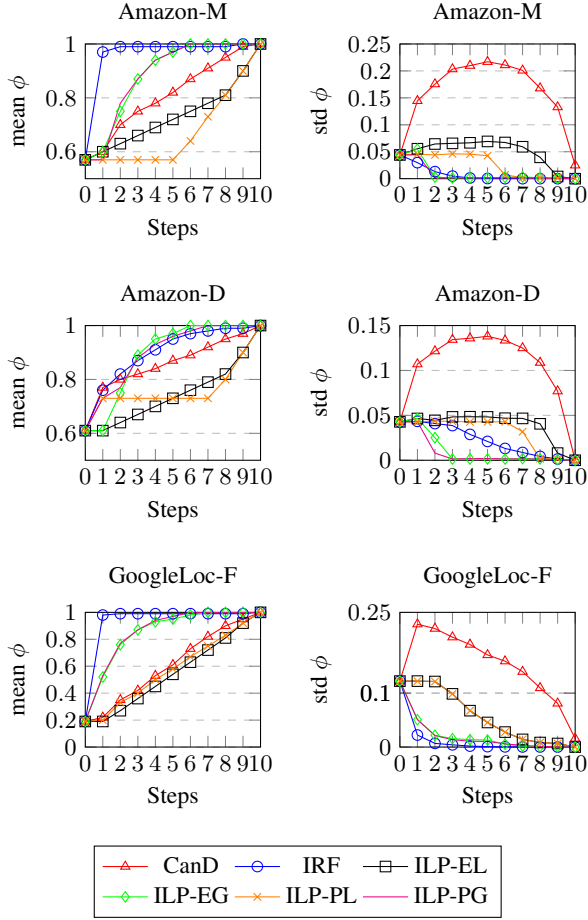
				
				\subsection{Producer-Side Results}
				Table~\ref{table:producer_metrics} reports the above metrics for all the baselines and proposed methods on different datasets.  
				
				\noindent \textbf{\textit{Performance of Baselines:}}
				{\em CanD} 
				ensures small ($\Pi$), performs very well in maintaining similar level of changes at each step ($Z$);
				but, {\em CanD} is less efficient due to high path lengths ($\Upsilon$) as the changes (change from $D^{i-1}$ to $D^{i}$ in step $i$) it introduces may or may not be directed towards $D^\eta$.
				The performance of {\em IRF} is not stable; in Amazon-M and Google-F, it performs very poorly in both the fairness metrics $\Pi$ and $Z$
				(the reason becomes clear when we look at  the customer side results);
				It also shows poor efficiency ($\Upsilon$) like {\em CanD}.
				
				\noindent \textbf{\textit{Performance of ILP-Estimated}} (EL/EG): 
				ILP-EL shows the lowest $\Pi$ and highest $Z$ making it the most fair method for the producers;
				it is very efficient ($\Upsilon$) too;
				ILP-EL with {\em prefiltering}(PF) also performs better (in $\pi$ and $Z$) than baselines and other ILP settings;
				however it incurs a loss in efficiency $\Upsilon$ due to its filtered item space.
				ILP-EG performs worse than ILP-EL in all metrics.
				
				\noindent \textbf{\textit{Performance of ILP-Preserving}} (PL/PG):
				ILP-PL shows very efficient ($\Upsilon$) transitions;
				however it performs very poorly in $\pi$ and $Z$ which makes it even more unfair than {\em CanD}.
				On the other hand, ILP-PG shows good results comparable to ILP-EL and ILP-EG.
				The reasons become clear when we study the individual $EC$ plots (elaborated next).
				
				\noindent $\pmb{EC}$ \textbf{\textit{Exposure Change Plots:}}
				We plot exposure changes ($EC$) at each of the steps of updates for Amazon-M 
				in Figure~\ref{fig:exp_change_amazon1}
				(similar plots for other datasets are provided in supplementary material).
				We see small and equal sized changes for {\em CanD} and ILP-EL, 
				however CanD produces slightly larger changes;
				this explains their similar $Z$ performances, 
				but different $\Upsilon$ and $\pi$.
				Both ILP-EG and ILP-PG show dissimilar changes in different steps;
				thus slightly larger $\Upsilon$ and $\pi$, and slightly lower $Z$.
				{\em IRF} causes huge change in the first step while ILP-PL shows large changes in the last few steps;
				this explains their high $\pi$ and very low $Z$ values.
				
				\subsection{Customer-Centric Metrics}
				In each step $i$ for each customer $u \in U([\tau_{i-1},\tau_i))$, 
				we obtain the utility $\phi^\text{norm}(R_{u'},V^\text{new})$;
				We calculate their mean and standard deviation and plot them in Figure~\ref{fig:utility_main}.
				The faster the mean utility grows, the faster the update applies.
				The standard deviation indicates the degree of unfairness.
				
				\subsection{Customer-Side Results}
				We explain the salient points of the results (in Figure~\ref{fig:utility_main}).\\
				\noindent \textbf{\textit{Performance of Baselines:}}
				The rise in mean customer utilities for {\em CanD} is comparable to ILPs, however for {\em IRF} it is much faster.
				Note that {\em CanD} incurs large standard deviation, i.e., it introduces larger disparity in customer utilities, which is undesirable.
				The {\em IRF} shows a large increase in mean utility in the first step of Amazon-M and GoogleLoc-F which essentially means it fails to update incrementally; 
				thus the producer fairness is severely compromised (correspondingly refer to Figure~\ref{fig:utility_main} and Table~\ref{table:producer_metrics}); 
				As in {\em IRF}, we choose the top-$k$ results using {intermediate relevance functions}, the intermediate function ($V^1$) at step $1$ drastically changes (it could have happened at any other step too) the top-$k$ set making it close to the top-$k$ of $V^\text{new}$;
				This explains the large increase in mean utility and the large exposure change in the first step in those datasets (refer Figure~\ref{fig:utility_main},\ref{fig:exp_change_amazon1} respectively).
				However, this is a very data-specific phenomenon as it doesn't happen in the Amazon-D.
				For the baseline methods, we see the above issues in ensuring producer and customer fairness;
				Reliability is also a major concern.
				
				\noindent \textbf{\textit{Performance of ILPs:}}
				By design, all ILPs ensure minimum utility guarantee to the customers in each of the steps. 
				The ILPs (EG and PG) with geometric steps in $\theta_i$ increases the 
				customer utilities quickly while the ILPs (EL and PL) with linear steps in $\theta_i$ 
				show slower improvements.
				In Amazon-D, Amazon-M, Google-Loc,
				the status quo (period $0$) mean utilities are near to 0.57, 0.61 and 0.19 correspondingly.
				Thus for ILP-Preserving (PL/PG), when there are scopes ($\theta_i$ becomes more than status quo utility),
				the ILPs show an update;
				This explains why ILP-PL generally shows significant updates (increase in utility Fig-\ref{fig:utility_main} and change in exposure Fig-\ref{fig:exp_change_amazon1}) only after some initial steps;
				while ILP-PG shows updates earlier due to geometric increase in $\theta_i$ and performs better.
				However for ILP-Estimated (EL/EG), such issue never comes as they enforce estimated changes in exposure (by setting {\small$\overline{D^i}$}) along with increase in $\theta_i$ in each step.
				The standard deviation of all the ILPs are small; the ILP-(EL/PL) have slightly higher values.
				
				\noindent{\bf Summary:} ILP-EL performs best in terms of producer fairness;
				its performance in maintaining customer utility is as per design;
				however, as the name suggests ILP-EL requires an estimation of change in producer exposure apriori.
				Whereas, ILP-PG performs a bit inferior to ILP-EL in terms of producer fairness but much better than baselines;
				the increase in customer utility is faster than ILP-EL.
				Most importantly, it doesn't require any estimation of the exposure change for designing each step which makes it an attractive choice.
				However, 
				our aim has been to explore a whole range of possibilities, 
				and leave it to the designer to choose one as per their requirement and available resources.

\section{Conclusion}
\label{discussion}
In this paper, we identified the adverse impact on the producers due to immediate updates in recommendations in two-sided platforms, 
and proposed an innovative ILP-based incremental update mechanism to tackle it.
Extensive evaluations over multiple datasets and different types of updates 
show that our proposed approach not only allows smoother transition of producer exposures, 
but also guarantees a minimum customer utility in 
intermediate steps. 
In future, we plan to check the impact of updates in more complex settings, such as when the assumption of 
closed market (where neither new producers/customers enter the system nor the overall demand fluctuates) is relaxed. 
We also plan to consider {\it position/ranking bias} in customer attention.
\subsubsection{Acknowledgements}
This research was supported in part by a European Research Council (ERC) Advanced Grant for the project ``Foundations for Fair Social Computing", funded under the European Union’s Horizon 2020 Framework Programme (grant agreement no. 789373). G. K Patro is supported by TCS Research Fellowship.
\bibliographystyle{aaai}
\bibliography{Main}

\appendix
\vspace{-4mm}
\section{Appendix}
\vspace{-2mm}
\subsection{Linear and Geometric Steps in $\theta_i$}
\vspace{-2mm}
We plot the customer utility lower bounds $\theta_i$ for step $i$ of the proposed ILPs for two different settings in Figure-\ref{fig:types_of_steps}.
As the lower bound $\theta_i$ grows much faster in geometric steps than in linear steps, we see faster growth of geometric steps setting.
This is why we see faster increase in mean customer utility in ILP-EG/PG than in ILP-EL/PL (refer Figure-\ref{fig:utility_main}).
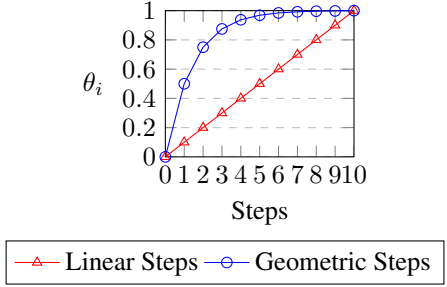
\begin{figure}[h!]
	\center{
		\subfloat{\pgfplotsset{width=0.23\textwidth,compat=1.9}
			\begin{tikzpicture}
			\begin{axis}[
			xlabel={Steps},
			ylabel={$\theta_i$},
			ylabel style={rotate=-90},
			xmin=0, xmax=10,
			ymin=0, ymax=1,
			xtick={0,1,2,3,4,5,6,7,8,9,10},
			ytick={0,0.2,0.4,0.6,0.8,1.0},
			ymajorgrids=true,
			grid style=dashed,
			]
			\addplot[
			color=red,
			mark=triangle,
			]
			coordinates {(0,0)
				(1, 0.1)
				(2, 0.2)
				(3, 0.3)
				(4, 0.4)
				(5, 0.5)
				(6, 0.6)
				(7, 0.7)
				(8, 0.8)
				(9, 0.9)
				(10, 1.0)
			};
			
			\addplot[
			color=blue,
			mark=o,
			]
			coordinates {(0,0)
				(1, 0.5)
				(2, 0.75)
				(3, 0.875)
				(4, 0.9375)
				(5, 0.96875)
				(6, 0.984375)
				(7, 0.9921875)
				(8, 0.99609375)
				(9, 0.998046875)
				(10, 1.0)
			};
			
			\end{axis}
			\end{tikzpicture}}
		\vspace{-3mm}
		\vfil
		\subfloat{\pgfplotsset{width=0.5\textwidth,compat=1.9}
			\begin{tikzpicture}
			\begin{customlegend}[legend entries={{Linear Steps},{Geometric Steps}},legend columns=2]
			\addlegendimage{red,mark=triangle,sharp plot}
			\addlegendimage{blue,mark=o,sharp plot}
			\end{customlegend}
			\end{tikzpicture}}	
	}\vspace{-3mm}
	\caption{Steps in $\theta_i$}
	\label{fig:types_of_steps}
	\vspace{-4mm}
\end{figure}
\vspace{-4mm}
\subsection{Consolidated $EC$ Plots}
\vspace{-2mm}
We plot $EC$ in each of the steps of all the baselines and proposed methods for all the datasets in Figure-\ref{fig:exp_change_all}.
The rows of plots are for certain methods and the columns are for certain datasets as titled in bold font.
The plots are generated with same setting as in the main paper (i.e., hyperparameters $k=10$, $\eta=10$).
\vspace{-4mm}
\subsection{Consolidated $\phi^\text{norm}$ Plots}
\vspace{-2mm}
In each step $i$ for each customer $u\in U([\tau_{i-1},\tau_i))$ (i.e., in period $i$), we obtain the utility $\phi^\text{norm}(R_{u},V^\text{new})$;
We calculate their mean, minimum and standard deviation and plot all of them in Figure-\ref{fig:utility_all}.
The faster the mean utility grows, the faster the update applies.
The standard deviation indicates the degree of unfairness.
The minimum utility (not discussed in main paper) in each step shows if any customer is too much deprived of utility in the corresponding step.
From the minimum utility plots, we see that our proposed ILPs have been able to maintain minimum customer utility more than the guaranteed lower bound.
However the baseline {\em CanD} has been unable to beat the same minimum utility;
this is because {\em CanD} chooses customers randomly in each step for whom it updates the recommendations;
those customers who have very less utility in status-quo and do not get selected for many number the steps, they continue to get very less utility.
\vspace{-4mm}
\subsection{Effects of $\eta$}
\vspace{-2mm}
We test the best performing (as discussed in \cref{experiments}) or the ILP with EL settings (estimated steps in {\small$\overline{D^i}$} and linear steps in $\theta_i$) with varying values for number of steps ($\eta$).
We find that by increasing the number of steps ($\eta$), we can ensure better transitions as illustrated in Figure-\ref{fig:tran_cost_steps} (transition cost decreases with increase in $\eta$).
However, we don't see any significant change in $\Upsilon$ and $Z$ in the same test.
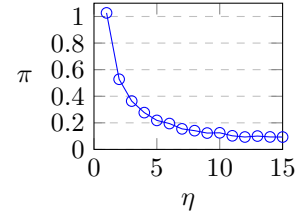
\begin{figure}[t!]
	\center{
		\subfloat{\pgfplotsset{width=0.23\textwidth,compat=1.9}
			\begin{tikzpicture}
			\begin{axis}[
			ylabel={$\pi$},
			xlabel={$\eta$},
			ylabel style={rotate=-90},
			xmin=0, xmax=15,
			ymin=0, ymax=1.1,
			xtick={0,5,10,15},
			ytick={0,0.2,0.4,0.6,0.8,1.0},
			ymajorgrids=true,
			grid style=dashed,
			]
			
			\addplot[
			color=blue,
			mark=o,
			]
			coordinates {(1, 1.0270109376858372)
				(2, 0.5279129601999231)
				(3, 0.3636822903056017)
				(4, 0.2769019794745468)
				(5, 0.2183289712235868)
				(6, 0.1947022122343939)
				(7, 0.15471084618716094)
				(8, 0.141110254228132)
				(9, 0.1230783180864536)
				(10, 0.12500331284161256)
				(11, 0.10252444137225823)
				(12, 0.09340249810387315)
				(13, 0.10053931349202055)
				(14, 0.09393561965473657)
				(15, 0.09314668727977943)
				
			};
			\end{axis}
			\end{tikzpicture}}	
	}
	\vspace{-4mm}
	\caption{Maximum transition cost $\pi$ vs. Number of steps $\eta$ (Amazon-M, ILP-EL)}
	\label{fig:tran_cost_steps}
	\vspace{-4mm}
\end{figure}
\vspace{-4mm}
\subsection{ILP at Producer-Level}
We extend our proposed ILP (as in Equation-\ref{eq:ilp_obj}) to producer-level as below.
{\small
	\begin{subequations}
		\begin{equation}
		\underset{X}{\text{argmin}} \mathlarger\sum\limits_{p\in P}\mathlarger\sum\limits_{s\in S_p}\bigg|E^t_p + \dfrac{X_{u',s}}{k} - (|U([\tau_{i-1},t))|+1)\cdot \overline{D^{i}_p}\bigg| \label{eq:ilp_obj_prod}
		\end{equation}
		s.t.
		\begin{equation}
		X_{u',s} \in \{0,1\} \forall s\in S \label{eq:ilp_const_1_prod}
		\end{equation}
		\begin{equation}
		\sum_{s\in S}X_{u',s}=k \label{eq:ilp_const_2_prod}
		\end{equation}
		\begin{equation}
		\sum_{s\in S}X_{u',s}\cdot V^\text{new}(u',s) \geq (\theta_i\cdot \phi^\text{max}_{u'}(V^\text{new})) \label{eq:ilp_const_3_prod}
		\end{equation}
	\end{subequations}}
Here the objective-\ref{eq:ilp_obj_prod} tries to minimize the sum of differences between observed exposure and targeted exposure of each producer while recommending items to $u'$ at time $t$ ($t\in [\tau_{i-1},\tau_i)$ or period $i$).
The constraints remain the same as in Equations-\ref{eq:ilp_const_1}, \ref{eq:ilp_const_2}, \ref{eq:ilp_const_3}.

We use the Amazon dataset and consider the first four characters of the item-id as the brand/producer of the corresponding item.
The resulting dataset has $16$ producers supplying $1-191$ items.
Next we test the above ILP with EL settings (estimated steps in {\small$\overline{D^i}$} and linear steps in $\theta_i$) on Amazon-M case with producer information;
we find the evaluation metrics as $\Upsilon=1.33$, $\pi=0.17$, and $Z=0.99$.
These values are very comparable to the item-level results of Table-\ref{table:producer_metrics}.

We also plot the $EC$ and mean $\phi^\text{norm}$ at each step in Figure-\ref{fig:Amazon1_prod_level}.
\begin{figure}[h!]
	\center{
	\subfloat[Amazon-m: $EC$ in each of the steps. Blue bars:$EC$ in steps and Red boldline: $EC$ for immediate update]{
		\begin{tikzpicture}
		\begin{axis}[
		xlabel=Steps,
		width=0.22\textwidth,
		height=0.18\textwidth,
		ylabel= $EC$,
		enlargelimits=0.05,
		ybar interval=0.61,
		xmin=0, xmax=11,
		ymin=0, ymax=0.7,
		xtick={2,4,6,8,10},
		ytick={0,0.3},
		extra y ticks = 0.663,
		extra y tick labels={0.663},
		extra y tick style={grid=major,major grid style={very thick,draw=red}}
		]
		\addplot[blue!20!black,fill=blue!80!red] 
		coordinates {(1, 0.0861)(2, 0.0803)(3, 0.0713)(4, 0.0739)(5, 0.0825)(6, 0.0994)(7, 0.1126)(8, 0.0874)(9, 0.0895)(10, 0.1009)(11,0)};
		\end{axis}
		\end{tikzpicture}}
		\hfil
		\subfloat[Mean $\phi$ in each of the steps]{
			\begin{tikzpicture}
			\begin{axis}[
			xlabel={Steps},
			ylabel={mean $\phi$},
			width=0.22\textwidth,
			height=0.18\textwidth,
			xmin=0, xmax=10,
			ymin=0.2, ymax=1,
			xtick={0,1,2,3,4,5,6,7,8,9,10},
			ytick={0,0.2,0.5,1},
			ymajorgrids=true,
			grid style=dashed,
			]
			
			\addplot[
			color=blue,
			mark=o,
			]
			coordinates {
				(0, 0.57)
				(1, 0.6042516015392131)
				(2, 0.6358359337762493)
				(3, 0.669572798422244)
				(4, 0.7024115883058737)
				(5, 0.7364788712728748)
				(6, 0.7693445103826941)
				(7, 0.8035480049068112)
				(8, 0.8428269714696524)
				(9, 0.9028608321294113)
				(10, 1.0)
			};
			
			\end{axis}
			\end{tikzpicture}}
	}\vspace{-3mm}
	\caption{Amazon-M Producer-Level ILP-EL}\label{fig:Amazon1_prod_level}
	\vspace{-4mm}
\end{figure}
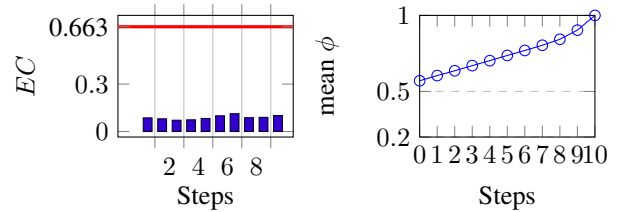
\vspace{-4mm}
\begin{figure*}[t!]
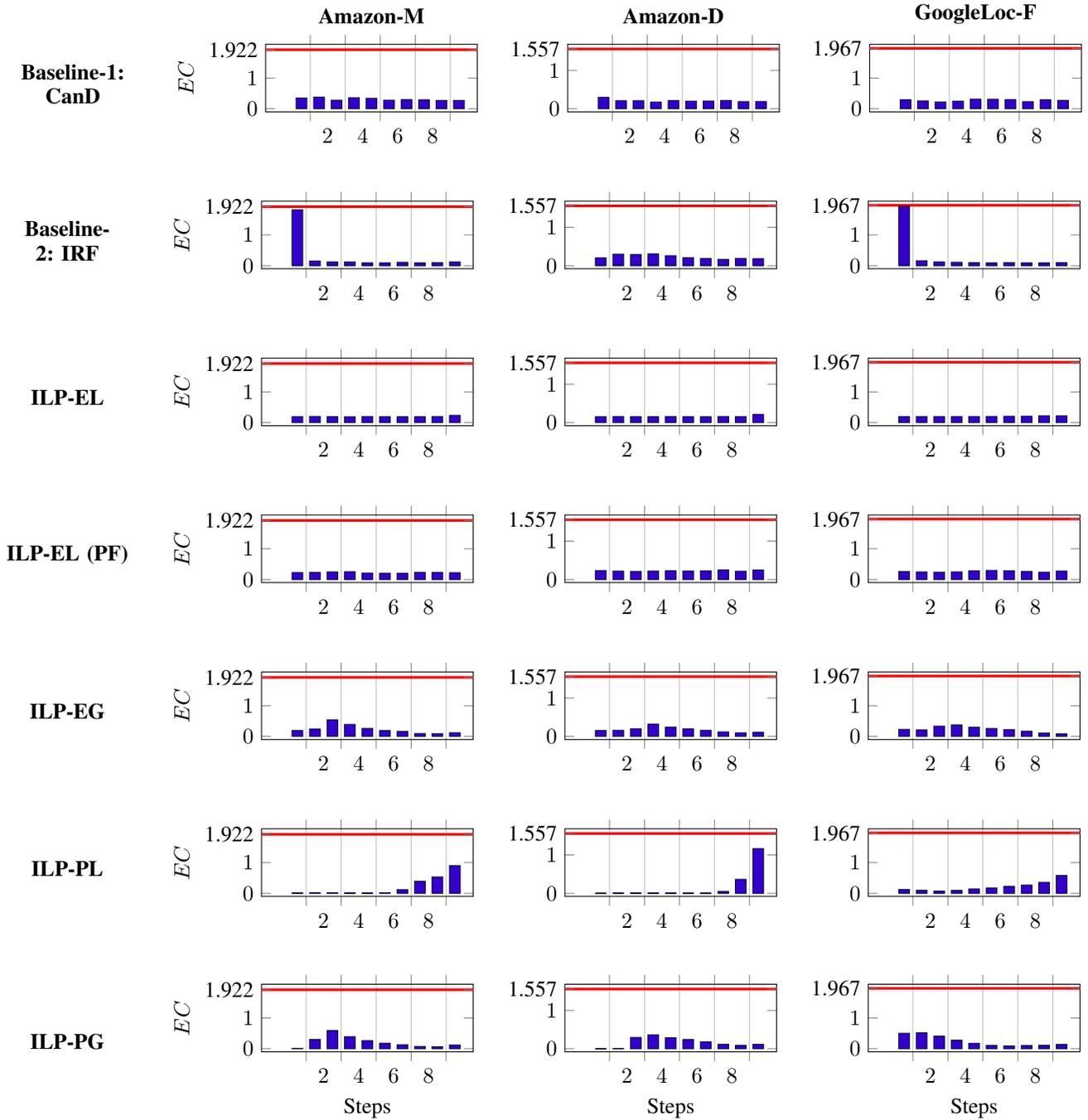

	\subfloat{\pgfplotsset{width=0.12\textwidth,compat=1.9}
}
	\caption{Exposure changes ($EC(i-1,i)=\sum_{s\in S}|D^i_s-D^{i-1}_s|$) in baselines and proposed ILPs (rows) on all the dataset (columns) are plotted (x-axis: steps $i$, y-axis: $EC$). The blue bars show the value of exposure change in each of the steps while the red boldline shows the exposure change caused by an immediate update. Hyperparameters: $\eta=10$, $k=10$
	}\label{fig:exp_change_all}
\end{figure*}

\begin{figure*}
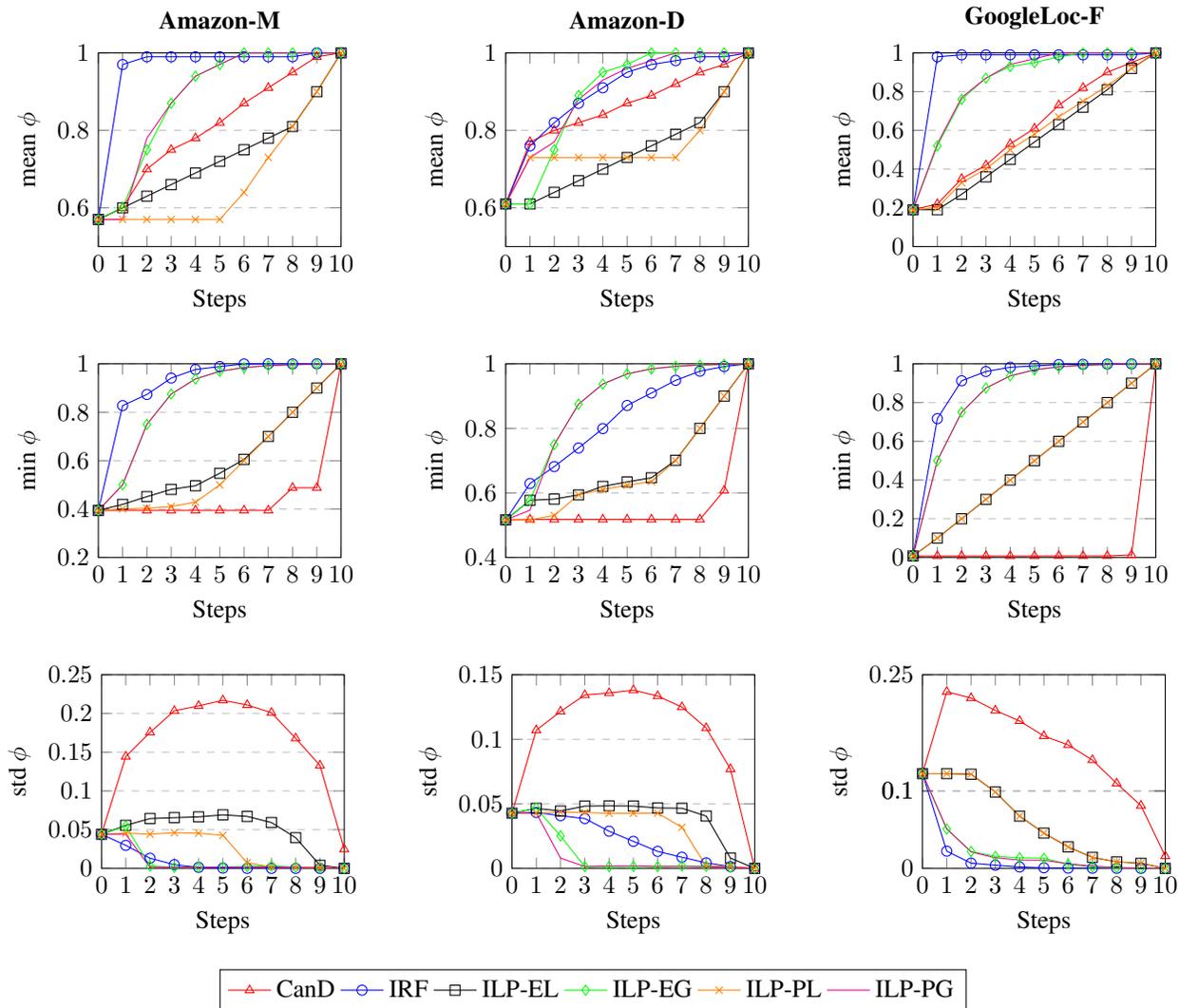

	\center{
		\subfloat{\pgfplotsset{width=0.28\textwidth,compat=1.9}
}
	}\vspace{-2mm}
	\caption{Mean (first row), minimum (second row) and standard deviation (third row) of $\phi^\text{norm}(R_u,V^\text{new})$ over all $u$ in each of the steps of the update are plotted; Amazon-M (first column), Amazon-D (second column), GoogleLoc-F (third column).}\label{fig:utility_all}
	\vspace{-2mm}
\end{figure*}

\end{document}